\documentclass[preprint,aps,nofootinbib,superscriptaddress,showpacs]{revtex4}
\usepackage{graphicx}
\usepackage{epsfig}
\usepackage{bm}
\usepackage{amssymb}

\newcommand{\lag}{\mathcal L}
\newcommand{\be}{\begin{eqnarray}}
\newcommand{\ee}{\end{eqnarray}}

\begin{document}
\title{Hyperon and nuclear symmetry energy in the neutron star}
\author{Chung-Yeol  Ryu}
\affiliation{Department of Physics,
Soongsil University, Seoul 156-743, Republic of Korea}
\author{Chang Ho Hyun}
\affiliation{Department of Physics Education, Daegu University,
Gyeongsan 712-714, Republic of Korea}
\author{Chang-Hwan Lee}
\affiliation{Department of Physics, Pusan National University,
Busan 609-735, Republic of Korea}

\date{July 27, 2011}

\begin{abstract}
In this work, masses and radii of neutron stars are
considered to investigate the effect of nuclear symmetry energy to
the astrophysical observables.
A relativistic mean field model with density-dependent
meson-baryon coupling constants is employed in describing the equation
of state of dense nuclear matter, and the
density dependencies of the symmetry energies are quoted from the recent
phenomenological formulae obtained from the heavy ion data at subnuclear
saturation densities.
Since hyperons can take part in the $\beta$-equilibrium of the dense matter
inside neutron stars, we include hyperons in our estimation
and their roles are discussed in combination with that of the nuclear
symmetry energy.
\end{abstract}
\maketitle

\section{Introduction}

Recent works about the nuclear symmetry energies at densities around the
nuclear saturation reveal an interesting diversity. $\pi^-/\pi^+$ production ratio in FOPI data of the heavy ion collisions at SIS/GSI favors super-soft
behavior of the symmetry energy,
which even becomes negative at densities above a few times the
saturation density \cite{Xia09}.
On the other hand, analysis of the giant monopole resonance for various nuclei,
and the determination of the neutron skin thickness show a monotonic increase of the nuclear symmetry energy
as density increases, even though the stiffness among them is substantially
different \cite{pieka07, cente09}.
Current status seems to demand more theoretical analysis as
well as ample experimental data in order to sharpen
our understanding on the nature of nuclear symmetry energy.

Neutron stars can provide a benchmark to test the density
dependence of the symmetry energy. Analysis for the neutron star with various forms
of symmetry energy was performed with a simple
free gas model \cite{kk09}, and the result for the matter composition in the interior of
the neutron star shows high sensitivity to the density dependence of the symmetry energy.
A softer symmetry energy allows more neutrons in the neutron star matter. On the other
hand a stiffer one requires more energy for the matter with higher asymmetry, and thus it favors
larger proton fraction.
With a very soft equation of state (EOS), it is difficult
for the neutron star matter to sustain the strong gravitational contraction,
and the star will easily collapse to a black hole.
Conversely, a stiffer EOS allows a larger value of the maximum mass of
a neutron star.
Well-measured neutron star masses lie in the range
of $(1.2\sim 2.0)M_\odot$ \cite{Lat07,Dem10}
and both super-soft and super-hard EOS which give neutron star masses far outside of this range can be excluded .
However, the maximum mass of neutron star is still an open question.

In this work, our main interest is the role of
symmetry energies which are fixed from the analysis of modern experiments
to the bulk properties of neutron stars.
It has been shown in numerous works that hyperons play a crucial
role in making the neutron star matter significantly soft and
consequently reduce the maximum mass of the neutron star
\cite{glendenning,prl90,npb91,prd95,prc95,plb95,prc96}.
When only nucleons are considered, soft symmetry energy will produce a matter exclusively
dominated by the neutron, but if the flavor changing $\beta$-equilibrium is allowed,
sufficiently high Fermi moment of the neutron makes the weak decay such as
$n \to \Lambda$ much more feasible.
These decays will have a critical impact on the composition, cooling rate,
and EOS of the neutron star matter.
Therefore it is essential to include hyperons in the neutron
star matter.

Production of hyperons and/or transition to exotic states
such as quark deconfinement or Bose-Einstein condensation are known
to be very sensitive to the input parameters such as coupling constants and
their density dependencies.
In Ref.~\cite{epja05}, we considered the density-dependent
masses and coupling constants for $\omega$ and $\rho$ mesons,
and showed that the density dependence affects the equation of
state and bulk properties of neutron stars critically.
In this work, we employ four empirical formulae for the symmetry
energy and plug them into a mean field model in which the coupling
constants are density-dependent.
By adopting such a hybrid method, we can single out the effect of
the symmetry energy on top of well-defined nuclear saturation
properties, and investigate its role to observables.

The empirical formulae for the symmetry energy are determined
from the data at subnuclear saturation densities in the relativistic
heavy ion experiments. Direct information about the behavior of the
symmetry energy at densities above the saturation is still lacking.
However, by extrapolating the existing formulae to supranuclear
densities, and calculating the neutron star properties with them,
we can indirectly constrain the behavior of the symmetry energy
at high-density region. The results indeed show that the different
behavior gives measurable differences in the properties of neutron
stars.


The paper is organized as follows. In Sec.~II, we introduce the models and formulae
used in the calculation. In Sec.~III, numerical results are displayed and we discuss them.
We conclude the paper in Sec.~IV.

\section{Model and Symmetry energy}

\subsection{Density-dependent relativistic mean-field model}
We briefly introduce the density-dependent relativistic mean-field (DDRMF)
model with octet baryons, and $\sigma$, $\omega$, and $\rho$ mesons
\cite{Len95,Fuc95,hkl01,tw99,BLi08}.
The model Lagrangian is given as
\be
\lag &=& \sum_B {\bar\psi}_B
\left[ \gamma_\mu
\left( i \partial^\mu - \Gamma_{\omega B} \omega^\mu
-\Gamma_{\rho B}\, {\vec b}^\mu\cdot \vec \frac{\tau}{2} \right)
- (m_B - \Gamma_{\sigma B} \sigma) \right ] \psi_B  \nonumber\\
&+& \frac 12\left(\partial_\mu\sigma \partial^\mu\sigma - m_\sigma^2\sigma^2\right)
- \frac 14 \omega_{\mu\nu}\omega^{\mu\nu} + \frac 12 m_\omega^2 \omega_\mu \omega^\mu
- \frac 14 {\vec b}_{\mu\nu}\cdot {\vec b}^{\mu\nu}
+ \frac 12 m_\rho^2 {\vec b}_\mu\cdot {\vec b}^\mu,
\ee
where $\Gamma_{iB}$ are density-dependent meson-baryon coupling constants, and
$\omega^{\mu\nu}(= \partial^\mu \omega^\nu - \partial^\nu \omega^\mu)$ and
${\vec b}^{\mu\nu}(=\partial^\mu {\vec b}^\nu - \partial^\nu {\vec b}^\mu)$
are the field tensors for $\omega$ and $\rho$ meson fields, respectively.

For the infinite nuclear matter, the equations of meson fields
in the mean-field approximation are obtained as
\be
m_\sigma^2 \sigma  &=& \sum_B \Gamma_{\sigma B}
\tilde{\rho}_B \nonumber\\
m_\omega^2 \omega_0  &=& \sum_B \Gamma_{\omega B} \rho_B \nonumber\\
m_\rho^2   b_{03} &=& \sum_B \Gamma_{\rho B} I_{B3} \rho_B,
\ee
where $\tilde{\rho}_B (= \bar{\psi}_B \psi_B)$ and
$\rho_B (= \psi^\dagger_B \psi_B)$ are the scalar and vector densities of
a baryon $B$, respectively, and $I_{B3}$ is its isospin $z$-component.
The Dirac equation of a baryon can be written as
\be
\left[\gamma_\mu \left(i\partial^\mu - \Sigma_\tau^\mu\right)
- \left(m_B - \Sigma_\tau^s\right)\right]\psi_B=0.
\label{eq-Dirac}\ee
If we assume $\Gamma_{iB} = \Gamma_{iB}(\rho)$ where $\rho = \sum_B \rho_B$,
the scalar and time component of vector self-energies
are obtained as
\be
\Sigma_\tau^s &=& \Gamma_{\sigma B}\, \sigma \nonumber\\
\Sigma_\tau^0 &=& \Gamma_{\omega B}\, \omega_0
+\Gamma_{\rho B} \, b_{03} I_{B3}
+ \Sigma_\tau^{0(r)}
\ee
with
\be
\Sigma_\tau^{0(r)} &=&
\sum_B \left [
\frac{\partial \Gamma_{\omega B}}{\partial \rho}\,\omega_0 \rho_B
+\frac{\partial \Gamma_{\rho B}}{\partial \rho} I_{B3} \rho_B b_{03}
-\frac{\partial \Gamma_{\sigma B}}{\partial \rho} \,\sigma \tilde{\rho}_B \right ].
\label{eq-rearr}\ee
Note that we only need time component of $\Sigma_\tau^\mu$ in the mean field approximation.

With the above equations, one can obtain the
energy density ($\varepsilon$) and the pressure ($P$) in the form
\be
\varepsilon  &=& \sum_B \frac{\gamma}{(2\pi)^3} \int_0^{k_F^B} d^3 k\,
\epsilon_B(k) + \frac 12 \left[m_\sigma^2 \,\sigma^2
+ m_\omega^2 \,\omega_0^2 + m_\rho^2 \,b_{03}^2 \right],
\nonumber\\
P &=& \sum_B \frac{\gamma}{3(2\pi)^3} \int_0^{k_F^B} d^3k
\frac{k^2}{\epsilon_B(k)}
+ \frac 12 \left[-m_\sigma^2 \,\sigma^2
+ m_\omega^2 \,\omega_0^2 + m_\rho^2 \,b_{03}^2 \right]
+ \rho \Sigma^{0(r)}_\tau, \label{eq-P}
\ee
where, $\epsilon_B(k)=\sqrt{ k^2 + (m_B^\star)^2}$ and $m_B^\star$ is the
effective mass of a baryon given by
$m_B^\star = m_B - \Gamma_{\sigma B} \,\sigma $.
Note that the rearrangement self-energy term $\Sigma^{0(r)}_\tau$
contributes to the pressure explicitly, but doesn't to
the energy density \cite{BLi08}.
Chemical potential of a baryon at its Fermi surface reads
\begin{eqnarray}
\mu_B = \frac{\partial \varepsilon}{\partial \rho_B} =
\epsilon_B(k^B_F) + \Gamma_{\omega B} \omega_0
+ I_{B3} \Gamma_{\rho B}\, b_{03} + \Sigma^{0(r)}_\tau.
\end{eqnarray}

\subsection{Coupling constants for symmetric matter}

In this work, we employ the density-dependent coupling constants
for the $\sigma$ and $\omega$ mesons proposed by
Typel \& Wolter (TW99) \cite{tw99}.
Coupling constants for the $\rho$ meson will be discussed in
the next subsection.
For $i = \sigma,\, \omega$, we can write
\begin{eqnarray}
\Gamma_{i B} (\rho) = g_{i B} f_i(n),
\end{eqnarray}
where $n = \rho/\rho_0$ with $\rho_0$ the saturation density.
It is assumed that $f_i(1)=1$, so $g_{iB}$ denotes the
coupling constant at the saturation density.
Density-dependent part $f_i(n)$ is given by
\begin{eqnarray}
f_i(n) = a_i \frac{ 1 + b_i (n + d_i)^2}{1+c_i (n + d_i)^2}.
\label{eq:dd}
\end{eqnarray}
Details for the procedure to fix the parameters in $f_i(n)$
can be found in \cite{tw99},
and we simply display the values of the parameters in
Table~\ref{tab:sigma-omega}, and the resulting saturation properties
in the caption.
\begin{table}
\begin{center}
\begin{tabular}{lllllll} \hline
Meson($i$) & $m_i$(MeV) & $g_{i N}$ & $a_i$ & $b_i$ & $c_i$ & $d_i$ \\ \hline
$\sigma$ & 550 & 10.87854 & 1.365469 & 0.226061 & 0.409704 & 0.901995 \\ \hline
$\omega$ & 783 & 13.29015 & 1.402488 & 0.172577 & 0.344293 & 0.983955 \\ \hline
\end{tabular}
\end{center}
\caption{Parameters of the density-dependent coupling constants in Typel \& Wolter
\cite{tw99} fitted to the saturation density $\rho_0 = 0.153\, {\rm fm}^{-3}$,
binding energy per nucleon $16.247$ MeV, and the compression modulus
$K_0=240$ MeV.}
\label{tab:sigma-omega}
\end{table}

For the couplings between hyperons and $\omega$-meson,
we assume the quark counting rule, i.e.
$g_{\omega \Lambda}= g_{\omega \Sigma} = \frac{2}{3} g_{\omega N}$
for $\Lambda$ and $\Sigma$, and
$g_{\omega \Xi} = \frac{1}{3} g_{\omega N}$ for $\Xi$.
For the hyperon-$\sigma$ meson coupling constants,
we fix them to reproduce the optical potential values for $\Lambda$, $\Sigma$ and $\Xi$
at the saturation density, $-30$ MeV, $30$ MeV and $-15$ MeV, respectively, which are determined empirically.
Numerical results thus determined are
$g_{\sigma Y}/g_{\sigma N} = 0.627,\,\, 0.480,\,\, 0.313$ for
$Y = \Lambda$, $\Sigma$ and $\Xi$, respectively.
As for the density dependence of the $\Gamma_{iY}$, we assume the
same dependence as given by $\Gamma_{iN}$, Eq.~(\ref{eq:dd}) for
simplicity.
The density dependence and coupling constants of $\rho$ meson will
be discussed in the next subsection.

\subsection{Symmetry energy beyond nuclear matter density}

Energy per baryon in infinite
nuclear matter is conventionally defined as
\begin{eqnarray}
E(\rho,\, \delta) = E(\rho,\, 0) + E_{\rm sym}(\rho)\,\delta^2 + O(\delta^4),
\end{eqnarray}
where $\rho = \rho_n + \rho_p$ is the nucleon number density and
$\delta = (\rho_n - \rho_p)/\rho$.
From the above equation, we can define the symmetry energy as
\begin{eqnarray}
E_{\rm sym}(\rho) = E(\rho,\,\, 1) - E(\rho,\,\, 0).
\end{eqnarray}

With the DDRMF model described above,
the energy density in the nucleonic phase is obtained as
\begin{eqnarray}
\varepsilon =
\sum_{i = n,\, p} \frac{\gamma}{(2\pi)^3} \int^{k^i_F}_0 d^3 k \epsilon_N(k)
+ \frac{1}{2} \left[m^2_\sigma \sigma^2  + m^2_\omega \omega^2_0
+ m^2_\rho b_{03}^2 \right].
\end{eqnarray}
Since the $\sigma$ and the $\omega_0$ fields depend only on
the baryon number density $\rho$, we obtain
\begin{eqnarray}
\varepsilon_{\rm sym} &=& \varepsilon(\rho,\,\, \delta = 1) - \varepsilon(\rho,\,\, \delta=0)
\nonumber \\
&=& \Delta E_{\rm kin} + \frac{\Gamma^2_{\rho N}}{8 m^2_\rho} \rho^2.
\end{eqnarray}
If we divide the symmetry energy into kinetic and potential terms as
\begin{eqnarray}
E_{\rm sym} = T_{\rm sym} + V_{\rm sym},
\end{eqnarray}
the kinetic term reads
\begin{equation}
T_{\rm sym} = \frac{\Delta E_{\rm kin}}{\rho}
= \frac{(k^{N}_F)^2}{6 \sqrt{(k^{N}_F)^2 + (m^{\star}_N)^2}},
\label{eq:kinetic}
\end{equation}
where $k^N_F$ is the Fermi momentum of the nucleon in symmetric
nuclear matter.
The potential part is then written as
\begin{equation}
V_{\rm sym} =  \frac{\Gamma^2_{\rho N}}{8 m^2_\rho}\, \rho.
\end{equation}
In the conventional considerations, $\Gamma_{\rho N}$
is assumed to be density-independent, and its value is determined
from the empirical symmetry energy at the nuclear saturation
density, $E_{\rm sym} = 30 \sim 35$ MeV .
However, recent works about the symmetry energy show
that its density-dependence can be much more significant than
what has been understood so far.
Especially, the uncertainties in its high-density behavior range  from
super-soft to very hard ones.

Remaining part of this subsection is devoted to the description
of the density-dependence of the symmetry energies employed in this
work.

\begin{itemize}
\item
{\bf MDI}

Symmetry energy has been considered in various theoretical
frameworks such as two- and three-body nuclear forces,
Brueckner-Hartree-Fock formalism, relativistic mean field theories,
effective field theories and etc.
In Ref.~\cite{chen05}, the authors included momentum-dependent interaction (denoted as MDI)
in the isovector channel of the interactions and obtained the so called
MDI version of the symmetry energy.
The result is parameterized in the form
\begin{eqnarray}
E_{\rm sym}(n) = T_{\rm sym}
+ A(x) n + [18.6 - A(x)] n^{B(x)},
\label{eq:soft}
\end{eqnarray}
where $x$ is a parameter fixed from experimental data.
Behavior of the symmetry energy is sensitive to the parameter $x$,
and three representative values are $1$, $0$ and $-1$.
Among these values, production ratio $\pi^-/\pi^+$ in FOPI data is
reproduced well with $x=1$,
with which $A(x) \simeq 107$~MeV and $B(x) \simeq 1.25$ \cite{Xia09}.
The potential terms derived from the above form are related to the
one in the mean field model as
\begin{eqnarray}
 \frac{\Gamma^2_{\rho N} \rho_0}{8 m^2_\rho} n = A(x) n
+ [18.6 - A(x)] n^{B(x)}.
\label{eq-rhoN}\end{eqnarray}
Since $B(x)$ is different from 1, if we divide both sides with $n$,
$n^{B(x)- 1}$ remains density-dependent,
and consequently we have a density dependence for the coupling constant
$\Gamma_{\rho N}$.

\item
{\bf FSU \& NL3}

Symmetry energy can be expanded in powers of $(n-1)$.
At subnuclear saturation densities, symmetry energy can be
approximated to quadratic order, and it can be written in the form
\begin{eqnarray}
E_{\rm sym}(n) = J + \frac{1}{3} L (n-1) + \frac{1}{18} K_{\rm sym} (n-1)^2,
\label{eq:hard}
\end{eqnarray}
where $J$, $L$ and $K_{\rm sym}$ are the parameters either calculated
from theories or fixed from the measured properties of nuclei.
FSU \cite{fsu} and NL3 \cite{nl3}
indicate the parameter sets of non-linear QHD model.
In the NL3 model, cubic and quartic terms of the self interactions of
$\sigma$ mesons are added to the quadratic QHD model.
In the FSU model, non-zero contribution of a term proportional to
$\omega^\mu \omega_\mu \, \vec{b}^\mu \cdot \vec{b}_\mu$ is added to the
NL3 one.
Coupling constants and parameters of the model are fitted to the static
properties of several nuclei. Once the parameters of the model are
determined, one can calculate the parameters that enter the
approximate form of the symmetry energy in Eq.~(\ref{eq:hard}).
Results for the nuclear saturation properties and the parameters
for the symmetry energy in Eq.~(\ref{eq:hard}) are shown
in Table~\ref{tab:jkl}.
\begin{table}[tbp]
\begin{center}
\begin{tabular}{ccccccc}\hline
Model &\,\, $\rho_0$\,\, &\,\, $E/A$\,\, &\,\, $K_0$\,\, &
\,\, $J$\,\, &\,\, $L$\,\, &\,\, $K_{\rm sym}$\,\, \\ \hline
FSU &\,\, 0.148\,\, &\,\, $-16.30$\,\, &\,\, 230.0\,\, &\,\, 32.59\,\, &\,\, 60.5\,\, &\,\, $-51.3$\,\, \\
NL3 &\,\, 0.148\,\, &\,\, $-16.24$\,\, &\,\, 271.5\,\, &\,\, 37.29\,\, &\,\, 118.2\,\, &\,\, 100.9\,\, \\ \hline
\end{tabular}
\end{center}
\caption{Nuclear satuation density $\rho_0$ is in fm$^{-3}$,
and the binding energy per nucleon $E/A$, the compression modulus $K_0$,
and the parameters in the symmetry energy $J$, $L$ and $K_{\rm sym}$
are in MeV.}
\label{tab:jkl}
\end{table}
Data from isospin diffusion and giant monopole resonance in heavy
ion collisions give the values $L = 88 \pm 25$~MeV and
$K_{\rm sym} - 6 L = -500 \pm 50$ MeV.
Values of $L$ and $K_{\rm sym}$ from both FSU and NL3 are within or
close to the empirical range.
One has to note that the kinetic term is not included in Eq.~(\ref{eq:hard}).
As a result, the potential term in the FSU and NL3 models reads
\begin{eqnarray}
\frac{\Gamma^2_{\rho N} \rho_0}{8 m^2_\rho} n =
 J + \frac{1}{3} L (n-1) + \frac{1}{18} K_{\rm sym} (n-1)^2 - T_{\rm sym}.
\end{eqnarray}

\item
{\bf TW99}

In Ref.~\cite{tw99}, the density dependence of $\Gamma_{\rho N}$
is given as
\begin{eqnarray}
\Gamma_{\rho N} = g_{\rho N} \exp[-a_\rho (n-1)].
\end{eqnarray}
A calculation based on Dirac-Brueckner approach showed that
the $\rho N$ coupling constant becomes small in magnitude
at high densities \cite{jong1998}.
The model gives $g_{\rho N} = 7.32196$, $a_\rho = 0.515$
and the symmetry energy $E_{\rm sym} = 33.39$ MeV at the
saturation density.
\end{itemize}

\begin{figure}
\centering
\includegraphics[width=7.5cm]{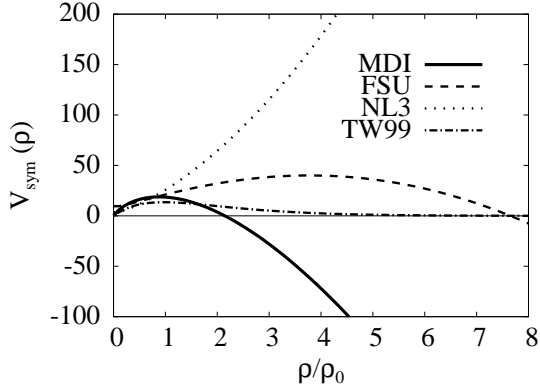}
\caption{$V_{\rm sym}(\rho) = \Gamma_{\rho N}^2 \rho/(8 m_{\rho}^2)$ for
four models.}
\label{fig:syme}
\end{figure}

In Fig.~\ref{fig:syme}, we show $V_{\rm sym}(\rho)$ for the four models.
In the MDI model, since $A(x) > 18.6$ and $B(x) > 1$, $V_{\rm sym}$ is an
increasing function at small $n$ and reaches maximum at $n = 0.88$.
After the maximum it decreases monotonically, and becomes negative
for $n> 2.1$.
A similar trend is observed from the curve for the FSU model.
%
%
%
%
In this model, the symmetry energy becomes negative for $n > 7.6$.

Since we assume $V_{\rm sym}(n) = \Gamma^2_{\rho N} \rho_0 n / 8 m^2_\rho$
in our calculation, negative values of $V_{\rm sym}$ imply negative values
of $\Gamma^2_{\rho N}$.
In the equations that determine the EOS, $\Gamma_{\rho N}$ always enters
in the quadratic form, so negative values of $\Gamma^2_{\rho N}$
cause no mathematical difficulty in solving the equations.
However, it can be problematic physically because a negative $\Gamma^2_{\rho N}$
value gives a pure imaginary number for the $\rho$-meson field, $b_{03}$.
We want to note that in most cases,
$V_{\rm sym}(\rho)$ of the models is determined
from the data at densities around or less than the saturation density.
It is not clear yet to what density the extrapolation is valid.
At the same time, it is not clear either to what density the term
$\Gamma^2_{\rho N} \rho_0 n/ 8 m^2_\rho$ will be dominant in the
symmetry energy. In other words, not only the quadratic term but also
higher order terms of the $\rho$-meson contribution in various forms
can be important at high densities.
Negative $\Gamma^2_{\rho N}$ value in the present work may be regarded
as either an unphysical result due to naive extrapolation of the
empirical formula to high densities, or a result of higher order
contributions
that can give rise to negative values of the symmetry energy.

In the NL3 model, $V_{\rm sym}$ is a quadratic function with a
positive coefficient for the quadratic term, and thus it is a
monotonically increasing function.
In the TW99 model, $\Gamma_{\rho N}$ decreases exponentially, so
$V_{\rm sym}$ converges to zero at high densities.

At $n>1$, the four models show very distinctive dependence on the
density, and we may classify NL3 to a hard model,
FSU and TW99 to a soft one, and the MDI to a super-soft one.
These controversial behaviors put distinguishing imprints on the
properties of neutron stars. In next section, we discuss the implications of symmetry energies to the properties of nuclear matter
inside neutron stars.

\section{Neutron Star Equation of State}

\begin{figure}[ht]
\centering
\includegraphics[width=7.5cm]{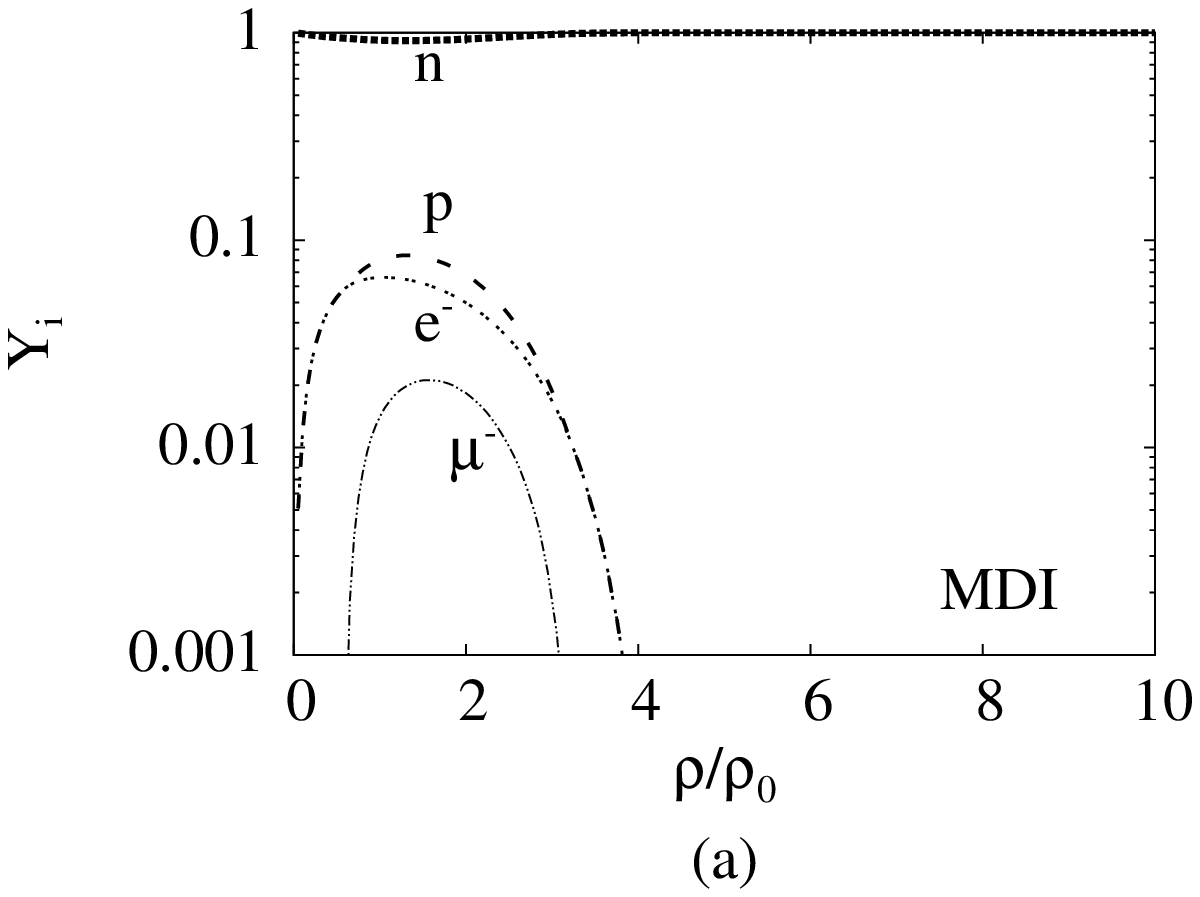}
\includegraphics[width=7.5cm]{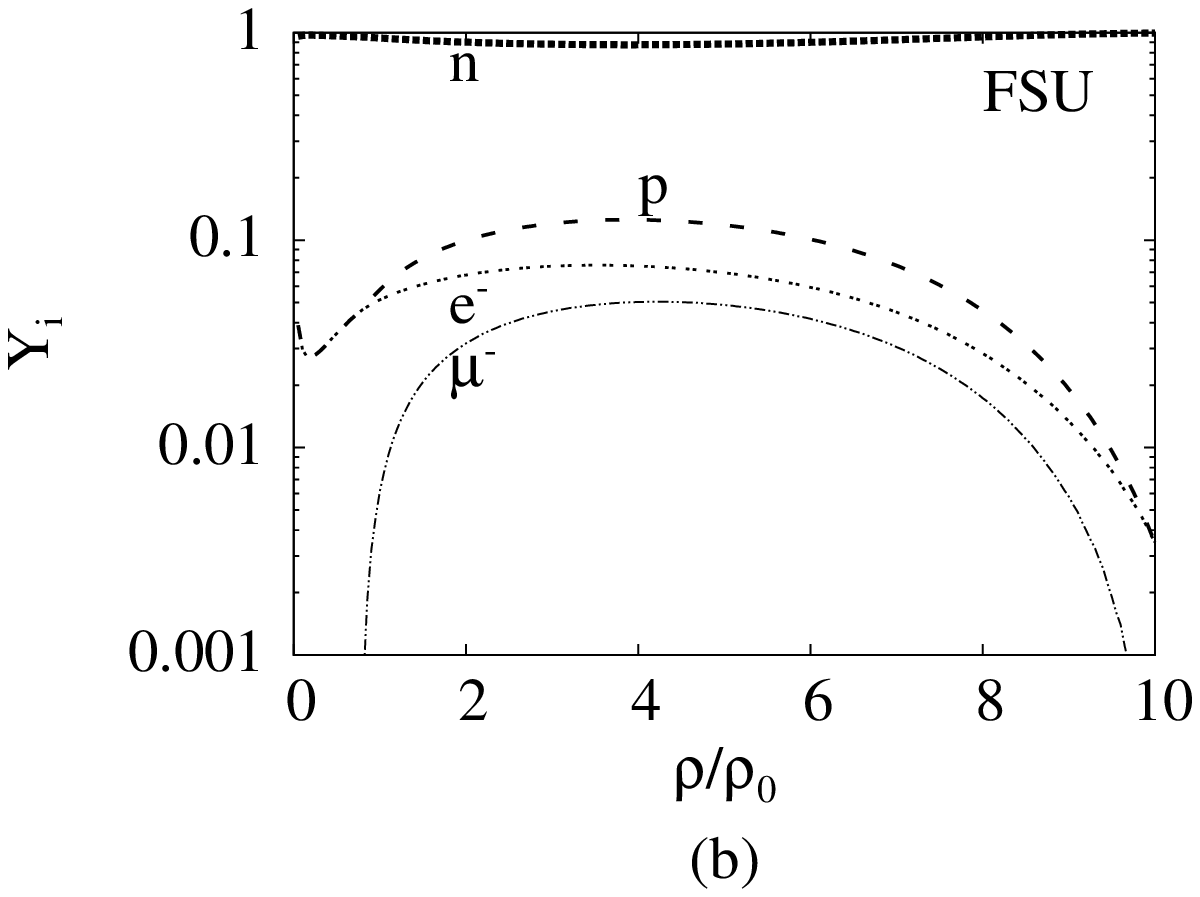} \\
\includegraphics[width=7.5cm]{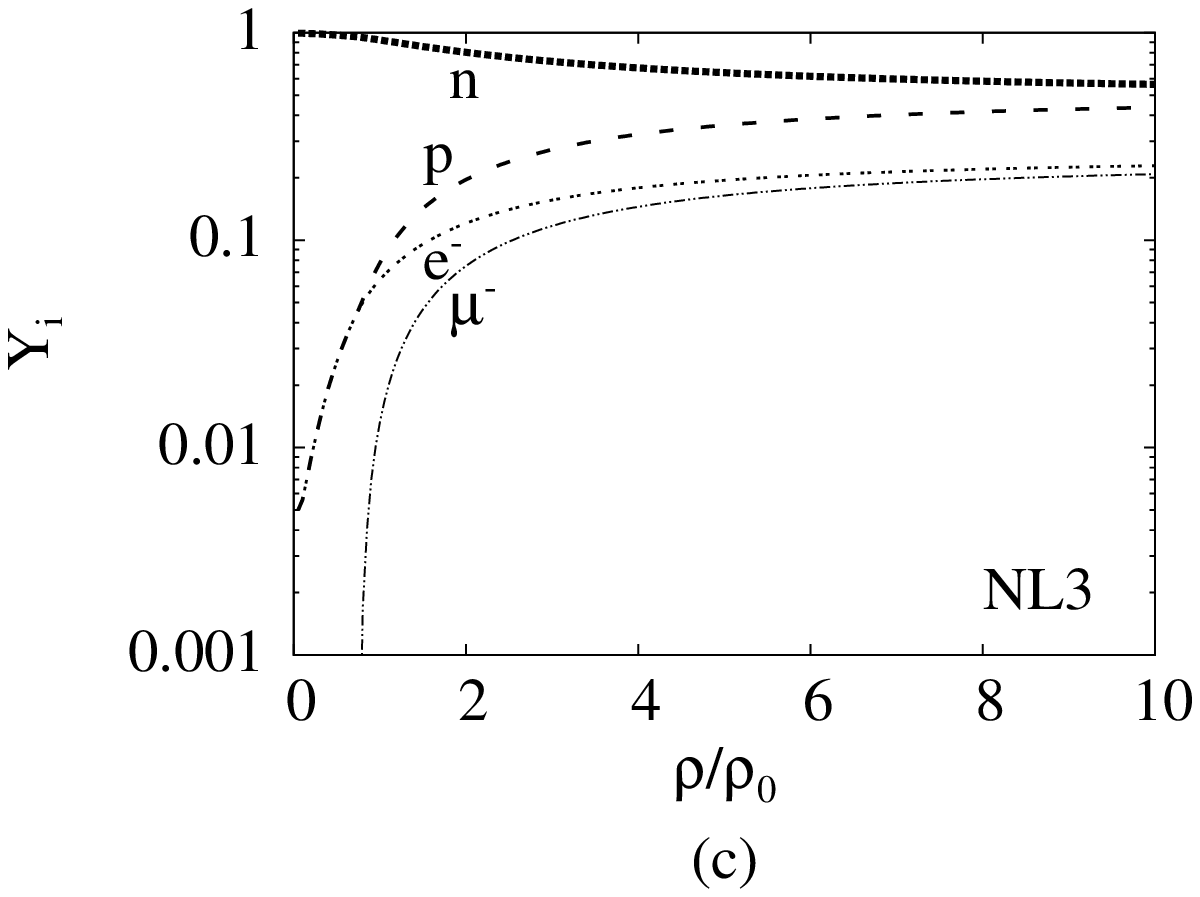}
\includegraphics[width=7.5cm]{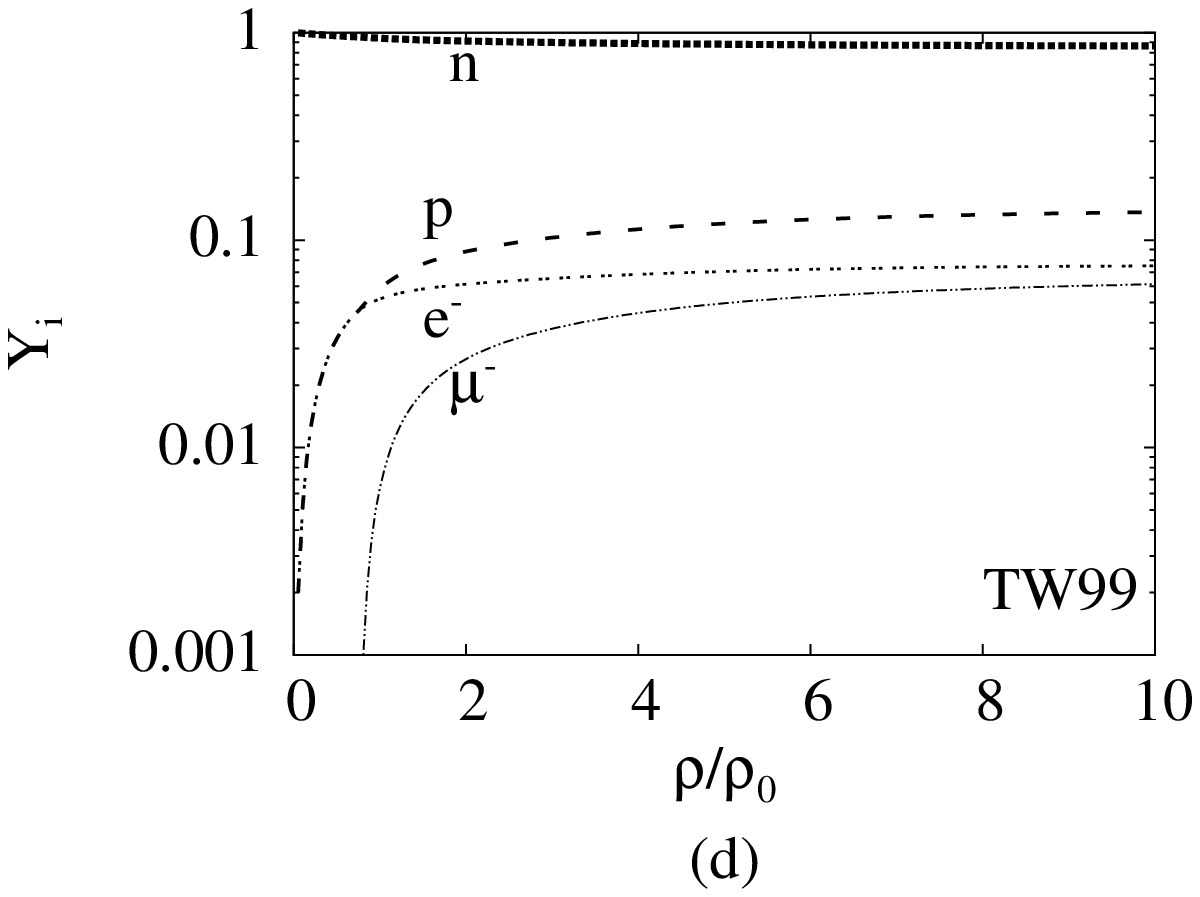}
\caption{Populations of particles for $np$.}
\label{fig:popul-np}
\end{figure}

In this section, we discuss the nuclear matter properties with various forms of symmetry energy. For the comparison we considered the nuclear matter with nucleons only ($np$) and with full baryon octet including hyperons ($npH$).

\subsection{Chemical composition of nuclear matter}
Figure~\ref{fig:popul-np} shows the fractional ratio of particle (nucleon and lepton) densities to the baryon density for $np$.
For the MDI model (Fig.~\ref{fig:popul-np}(a)), the proton fraction is
less than 10~\% at densities where $V_{\rm sym} > 0$.
As $V_{\rm sym}$ decreases quickly at high densities,
only the neutrons remain in the star.
Such an unusual particle composition can
be understood from the behavior of $V_{\rm sym}$:
Negative values of $V_{\rm sym}$ favor more asymmetric
matter than the symmetric one.
With the proton fraction less than 10~\%,
the direct URCA process, $n \to p e^- \bar{\nu}_e$
cannot happen in the star.
If there are only neutrons, then the weak decay
will no longer happen, and neutrino emission will cease
from the interior of the star.
Emission of the neutrino has a significant effect to
the thermal evolution of the neutron star, and thus,
though it is beyond the scope of this paper, it is
important to investigate whether the symmetry energies
are compatible with the cooling curve of the neutron star.
The particle fraction for FSU model (Fig.~\ref{fig:popul-np}(b)) shows similar
qualitative behavior as that of the MDI model in the region where the symmetry energies become negative.

In the TW99 model, $V_{\rm sym}$ converges to zero
at high densities, which means that the isospin dependent
interaction doesn't affect the $\beta$-equilibrium
condition any more. Since $\sigma$- and $\omega$-mediated
interactions are isospin-independent,
only the kinetic term determines the particle
fraction as $V_{\rm sym}$ approaches zero in the TW99 model.
This is the reason why the particle fractions saturate at high densities
in this model as one can see in Fig.~\ref{fig:popul-np}(d).
On the other hand, $V_{\rm sym}$ increases monotonically with density in
the NL3 model (Fig.~\ref{fig:popul-np}(c)).
As a result the proton fraction increases consistently as density increases.

\begin{figure}
\centering
\includegraphics[width=7.5cm]{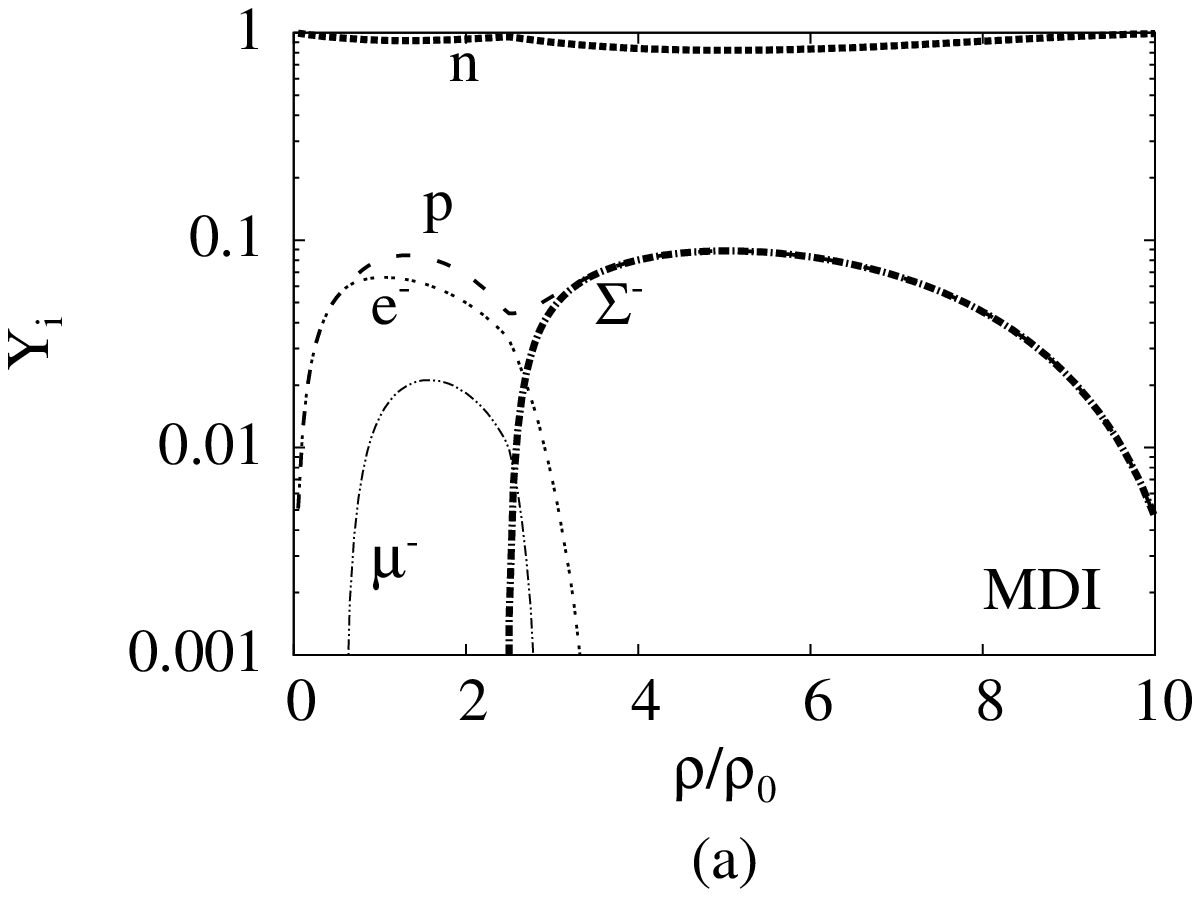}
\includegraphics[width=7.5cm]{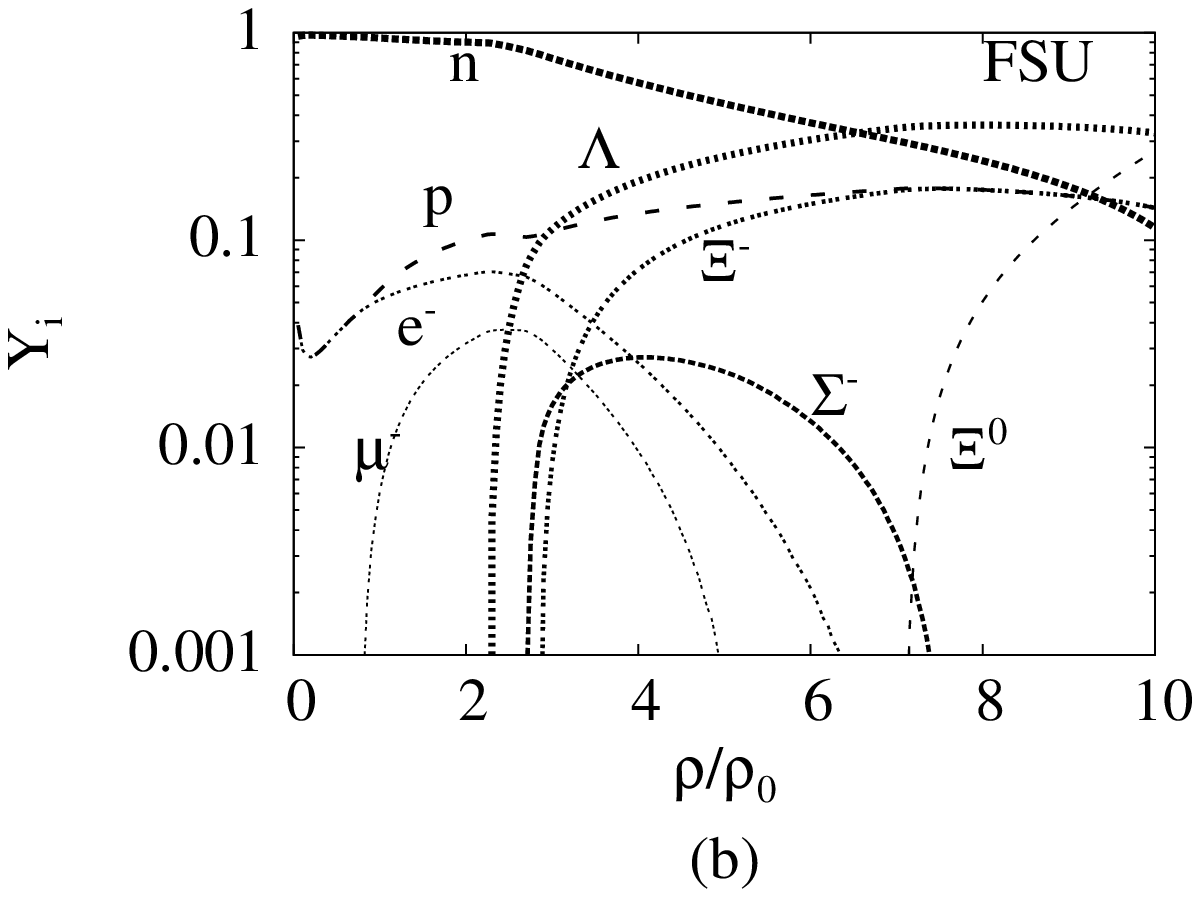} \\
\includegraphics[width=7.5cm]{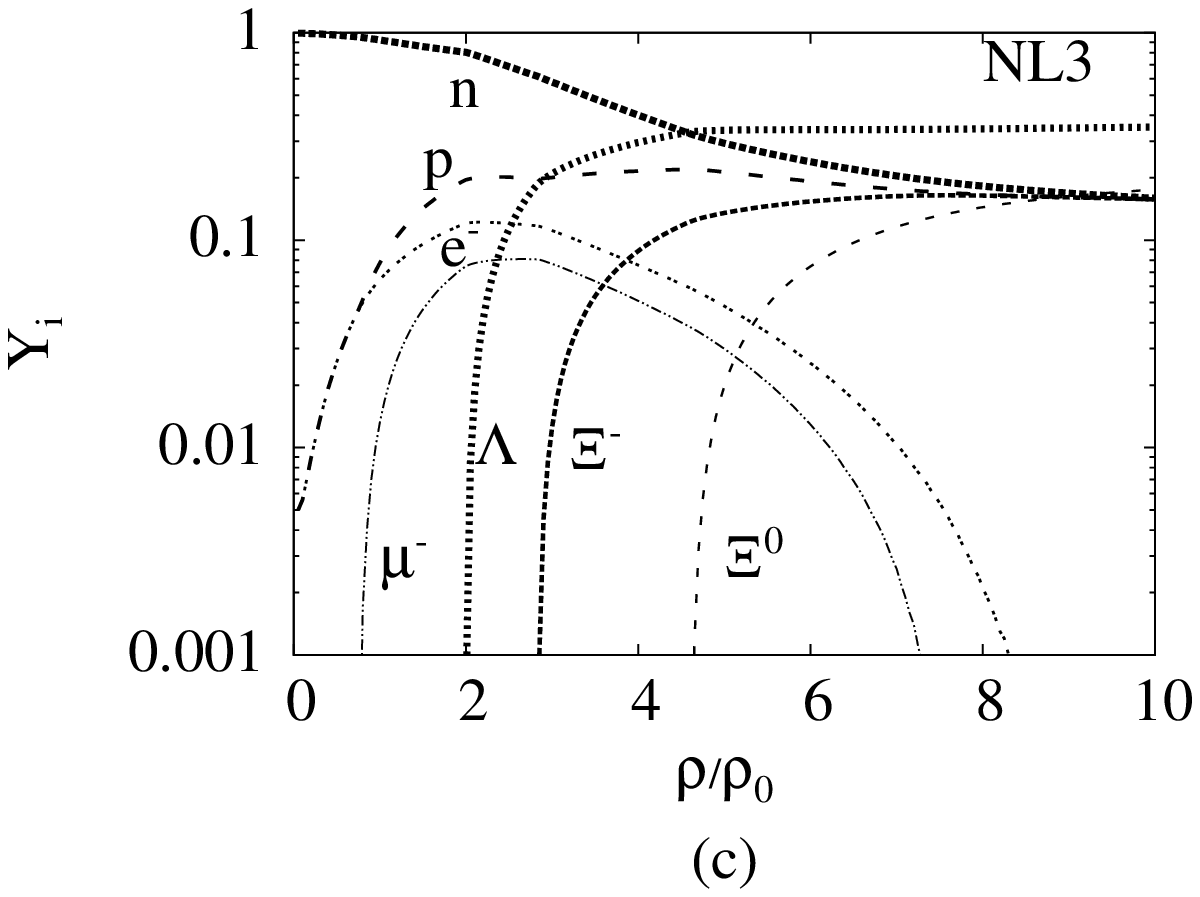}
\includegraphics[width=7.5cm]{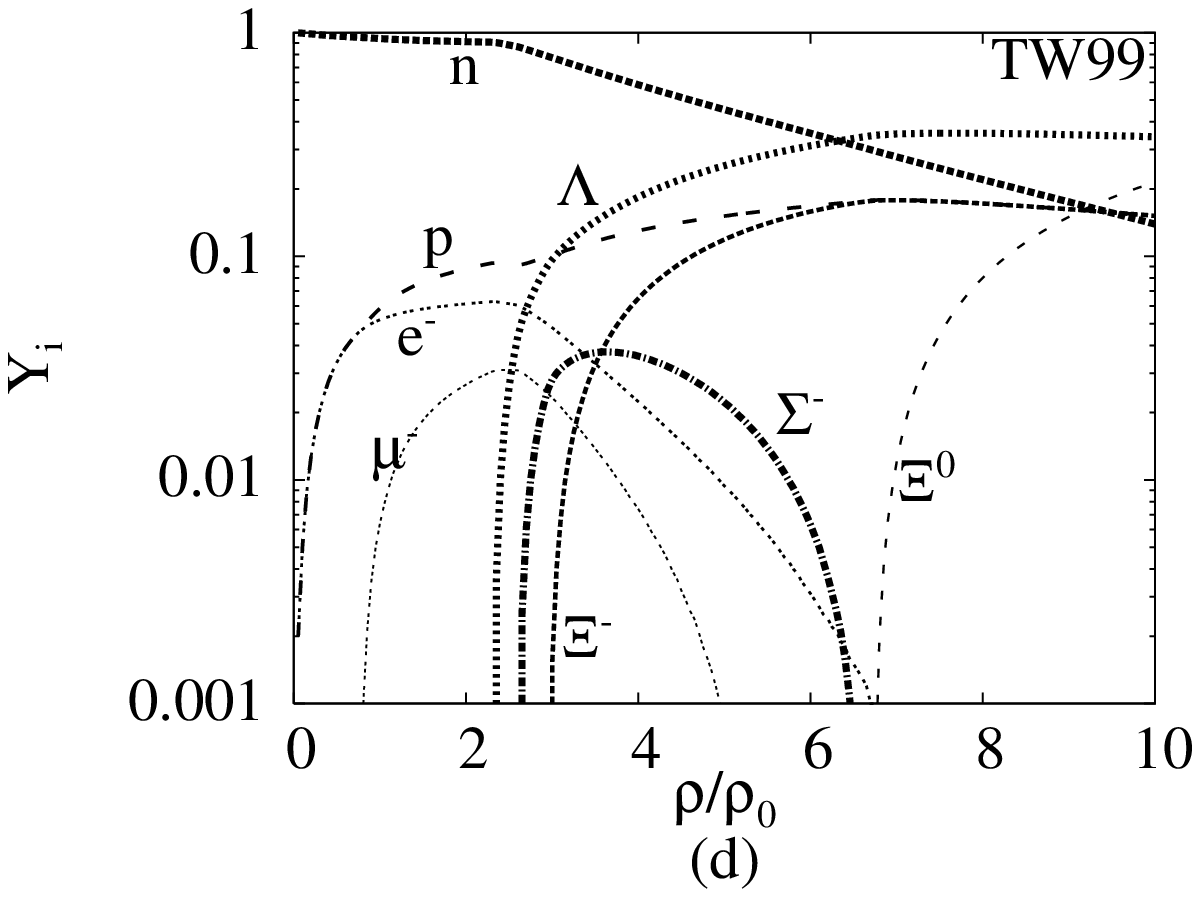}
\caption{Populations of particles for $npH$.}
\label{fig:popul}
\end{figure}

In Fig.~\ref{fig:popul}, we show the fractional ratio of particles for $npH$.
One can see that the critical densities for the creation of hyperons, and
their fractional ratios above the critical densities are very sensitive
to the density dependence of the symmetry energy.
With hyperons, an unusual result appears from
the MDI model (Fig.~\ref{fig:popul}(a)),
which is very contrasting to those of the remaining models.
$\Lambda$ is the lightest hyperon, and at the same time it
feels the strongest attraction among the hyperons in the baryon octet.
For these reasons, $\Lambda$ hyperon is the first hyperon that
are created via $\beta$-equilibrium condition
in the FSU, NL3 and TW3 models
(Fig.~\ref{fig:popul}(b), (c), (d), respectively).
On the other hand, since the $\Sigma$ hyperons feel the
strongest repulsion among the hyperons
in the nuclear matter, their chemical
potential is always too large to satisfy the
$\beta$-equilibrium conditions, so it never appears in the
NL3 model.
Similar result was obtained from the quark-meson coupling
model \cite{ryu07}.
In the result with MDI model, however, even though $\Sigma$ hyperon is
repulsive, it is the only hyperon that satisfies the
$\beta$-equilibrium condition.

We can understand the unusual behavior in the MDI model
from the chemical potentials.
The chemical potential of $n$, $\Lambda$ and $\Sigma^-$
read respectively as,
\begin{eqnarray}
\mu_n &=& \epsilon_n + \Gamma_{\omega N} \omega_0 +
\frac{1}{4} \left( \frac{\Gamma_{\rho N}}{m_\rho}\right)^2
(\rho_n - \rho_p), \\
\mu_\Lambda &=& \epsilon_\Lambda + \Gamma_{\omega \Lambda} \omega_0, \\
\mu_{\Sigma^-} &=& \epsilon_{\Sigma^-} + \Gamma_{\omega \Sigma} \omega_0 +
\frac{1}{2} \frac{\Gamma_{\rho N}\Gamma_{\rho \Sigma}}{m^2_\rho}
(\rho_n - \rho_p).
\end{eqnarray}
In the MDI model, $\rho_n - \rho_p$ increases in the region where
$V_{\rm sym}$ decreases.
Even when $V_{\rm sym} < 0$, $\rho_n-\rho_p$ keeps increasing,
but at densities where $V_{\rm sym}$ is negative,
the $\rho$-meson contribution to $\mu_n$ becomes
negative. In other words, in the NL3 model, the $\rho$-meson
contribution to $\mu_n$ is always positive but it becomes negative
above a certain density in the MDI model.
On the other hand, since there is no $\rho$-meson contribution to
$\mu_\Lambda$, the chemical potential of $\Lambda$ hyperon increases
with density.
If the negative contribution of $\rho$-meson to $\mu_n$ becomes
significant before the $\beta$-equilibrium condition $\mu_n=\mu_\Lambda$
is satisfied, then there is no chance to have $\Lambda$ hyperons in
the interior of the neutron star.
As for the $\Sigma$ hyperon, since $\Gamma_{\rho \Sigma} \propto \Gamma_{\rho N}$ in our calculation,
when the $\rho$-meson contribution to $\mu_n$ is negative, it is also
negative to $\mu_{\Sigma^-}$.
Since the $\beta$-equilibrium for $\Sigma^-$ reads
$\mu_n + \mu_{e^-} = \mu_{\Sigma^-}$, if the Fermi momentum of electrons
increases sufficiently to satisfy the $\beta$-equilibrium
condition for $\Sigma^-$, we can
have $\Sigma^-$ prior to $\Lambda$, as summarized in Fig.~\ref{fig:popul}.

An early creation of $\Lambda$ hyperon in the NL3 model than in the
FSU and TW99 models is also interesting even though the differences in the critical densities are not so big.
The $\beta$-equilibrium condition $\mu_\Lambda = \mu_p+ \mu_{e^-}$,
and the behavior of symmetry energies
provide us a simple and useful insight into the origin of the difference.
Because $V_{\rm sym}$ in the NL3 model is stiffer than those in
the FSU and TW99 models,
the proton fraction in the NL3 model is roughly twice of those in the
FSU and TW99 models at densities around $n=2$.
A larger proton fraction gives a larger Fermi momentum of the proton and the
electron, hence the early creation of $\Lambda$ is favored, reducing the energy of the system.
%

\subsection{Neutron star equation of state}

\begin{figure}
\centering
\includegraphics[width=7.5cm]{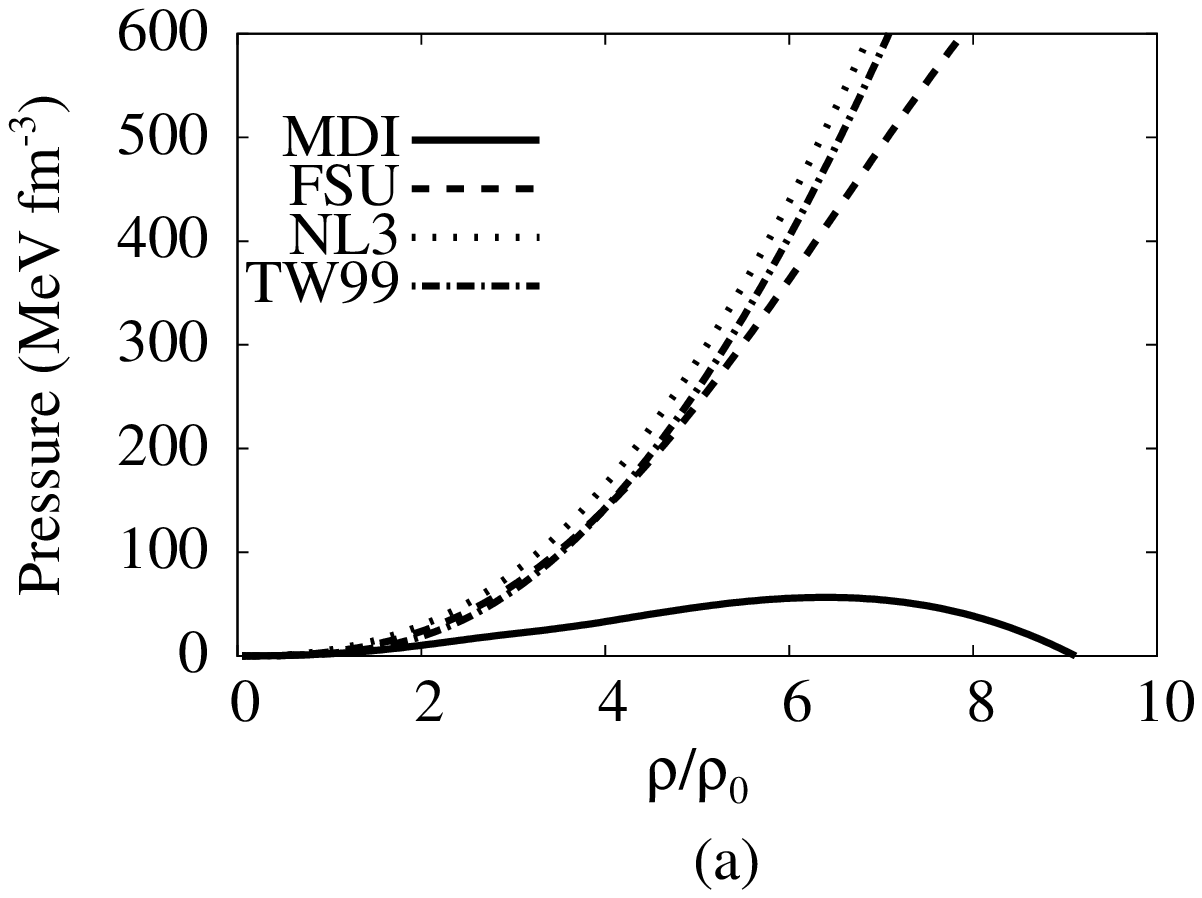}
\includegraphics[width=7.5cm]{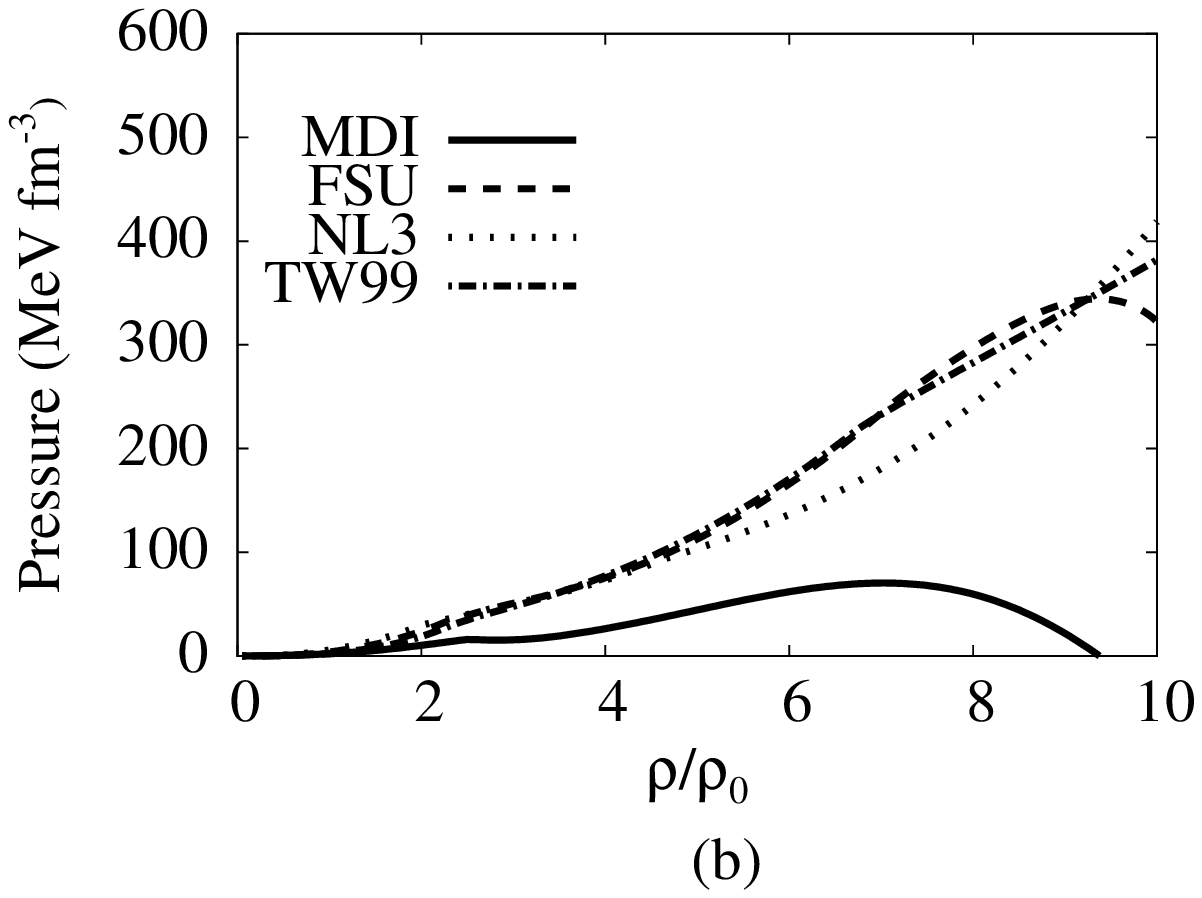}
\caption{Equation of state for $np$ (left pannel) and $npH$ (right pannel).}
\label{fig:eos}
\end{figure}

In Fig.~\ref{fig:eos}, we summarize the pressure as a function
of density for both $np$ (Fig.~\ref{fig:eos}(a)) and
$npH$ (Fig.~\ref{fig:eos}(b)).
Since we use a single model for the isospin independent interactions,
the difference in the pressure directly reflects the different
contribution from isospin dependent interactions,
$V_{\rm sym}$ as in Eq.~(\ref{eq-P}).
Similar to the curves for $V_{\rm sym}$ in Fig.~\ref{fig:syme},
the NL3 model exhibits the stiffest increase among the models,
the FSU and TW99 models show milder behavior than the NL3,
and the MDI model shows dramatic softening of the EOS due to the
symmetry energy.
In the MDI model, the pressure decreases above $\rho = 6.5 \rho_0$,
which is a signal for the instability.
In the results with hyperons, however, the curves are more complicated.
With a stiffer symmetry energy, we discussed that hyperons are created
at lower densities.
Because the isospin-independent repulsive interaction of hyperons is
weaker than that of the nucleon, the appearance of hyperons makes the
pressure softer in the isospin-independent part.
Complicatedness of the curves for the pressure with hyperons than
that with only nucleons may be due to the competition between
the enhancement from the isospin-dependent interaction and
the reduction from the isospin-independent one.
Comparing the pressure from NL3 to the ones from FSU or TW99,
one can see that the pressure from NL3 is stiffer at low densities,
but it becomes softer at high densities.
This behavior is consistent to the result in Ref.~\cite{chen10}.
Regardless of the density-dependence of the symmetry energy,
it is certain that the EOS of the neutron star matter is dramatically
softened by the creation of the hyperons.
Effect of the symmetry energy
can be explored by calculating the mass
and the radius of the neutron star, and comparing them with available
observations.

\begin{figure}
\centering
\includegraphics[width=7.5cm]{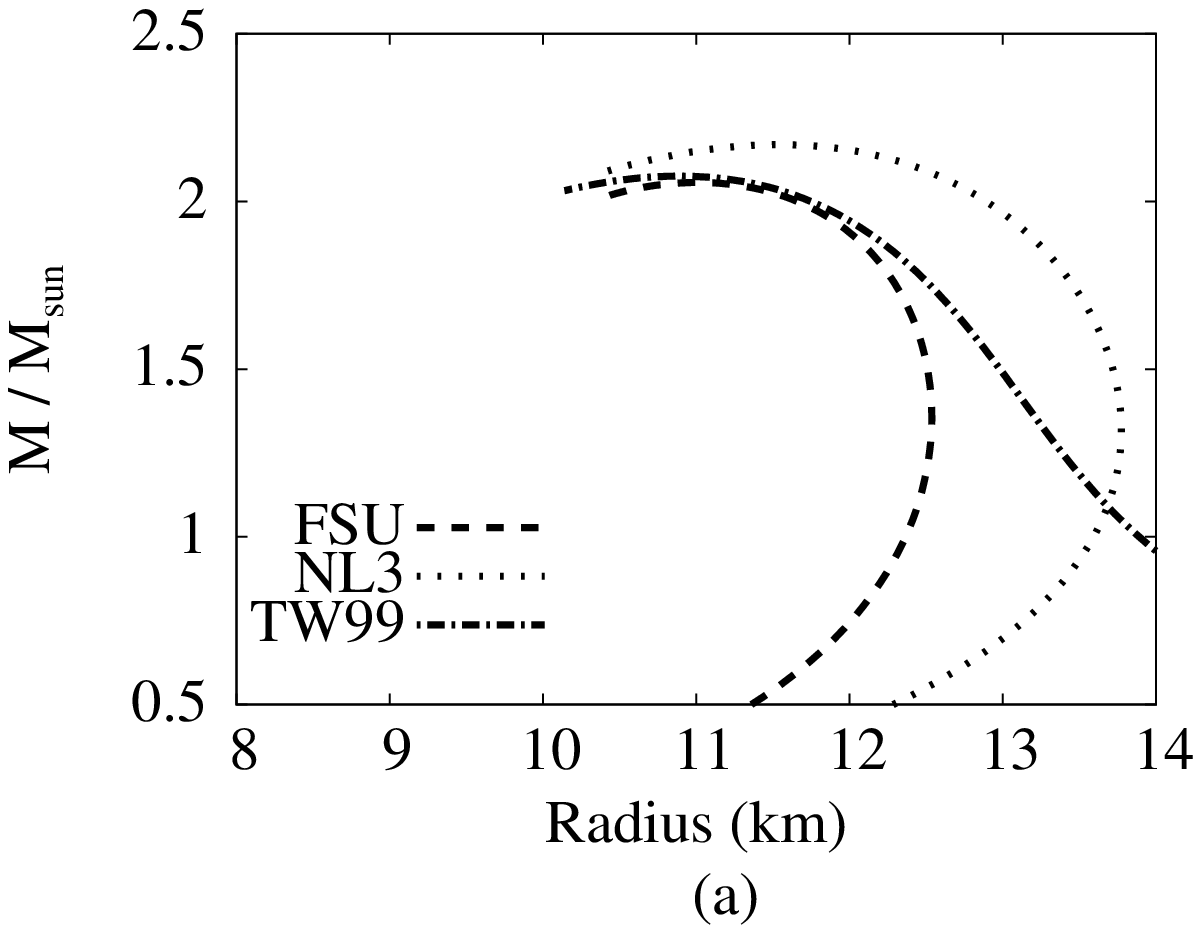}
\includegraphics[width=7.5cm]{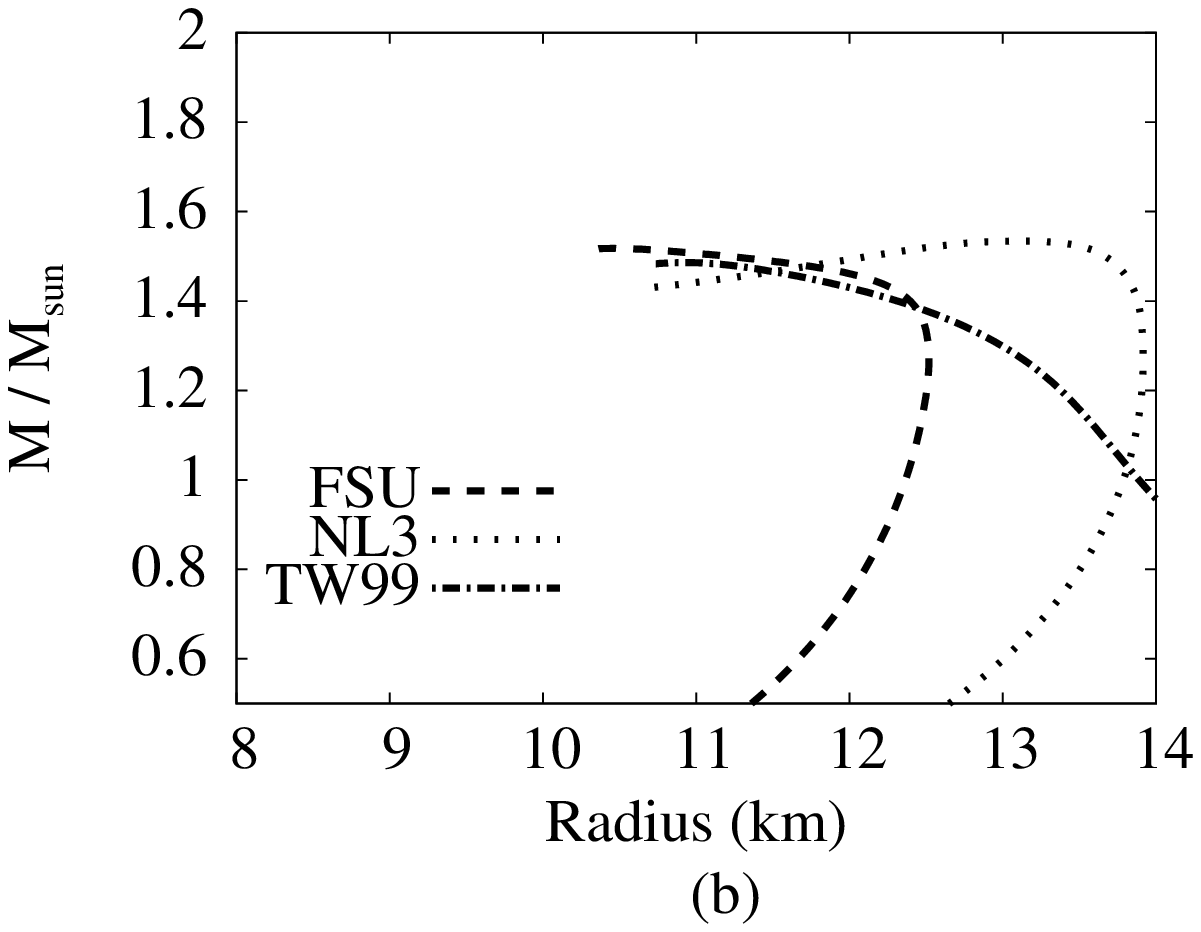}
\caption{The relation between mass and radius for $np$ (left pannel) and $npH$ (right pannel).}
\label{fig:mr}
\end{figure}

Figure~\ref{fig:mr} shows the mass-radius relation of the neutron
star for $np$ (Fig.~\ref{fig:mr}(a)) and $npH$ (Fig.~\ref{fig:mr}(b)).
In Fig.~\ref{fig:mr}(a), NL3 shows the largest mass $\sim 2.2 M_\odot$
which is well expected from its stiffest EOS. The difference from the
other models is, however, not so much, approximately $0.1 M_\odot$.
The maximum mass around $2 M_\odot$ is a standard value
which one can find in many works with the mean field theories.
With octet baryons, the maximum mass drops to $\sim 1.6 M_\odot$
(Fig.~\ref{fig:mr}(b)).
The difference in the maximum mass among the models is even smaller
than that with the nucleons only, but the radius of the star at the maximum mass exhibits
noticeable difference.
For the FSU and the TW99 model, the radius at the maximum mass
is in the range 10.5 - 10.9 km, but the radius from the
NL3 model is about 13.1 km.
Precise measurement of the neutron star radius can give
meaningful information and constraint to the nuclear
symmetry energy at high densities.
With the MDI model, on the other hand,
we couldn't obtain a stable configuration
of the star.
It may imply that such a super-soft symmetry energy obtained from
extrapolation from low densities is unrealistic at high densities,
and one has to have more direct information from high-density region.

\section{Discussion}


In this work, we estimated the contribution of symmetry energy to the EOS
of neutron stars with four different sets of parameterization.
We also estimated the contribution of hyperons by employing
DDRMF model without ($np$) and with hyperons ($npH$).
In both $np$ and $npH$ cases, with soft symmetry energy,
the neutron fraction increases as density increases at high densities,
causing highly asymmetric dense nuclear matter.
On the other hand, with hard symmetry energy, the neutron fraction
decreases, and it makes the matter more symmetric.
The dramatic change in the composition of particles in neutron star
matter due to the density dependence of symmetry energy is
very important to understand both the experimental results of
heavy ion collisions and the properties of neutron stars.
In other words, heavy ion experiments and neutron stars can give
constraints on the symmetry energy and the EOS of dense matter.

Recent analysis on the $\pi^-/\pi^+$ ratio in FOPI data of GSI
heavy ion experiments \cite{Xia09} raised questions on the behavior
of symmetry energy at high densities. They showed that very soft
symmetry energy is favored to explain the experimental result.
However, since the symmetry energy becomes negative at high densities,
which is contradictory to the naive expectation based on the QCD symmetry,
it still requires more careful investigation.

Other possibilities which we didn't consider in this work is
the phase transition to exotic states
such as boson condensation or transition to quark matter.
Since kaon is the lightest boson with strange quark, and kaons will
compete with hyperons in the $\beta$-equilibrium condition,
kaon condensation at high densities can be important.
The work with kaon condensation is in preparation.

We confirmed that the soft EOS reduces the maximum mass of neutron stars.
However, with the super-soft EOS in the MDI model, stable neutron star
cannot be constructed. This indicates that the super-soft EOS may not be
realistic as long as we accept the standard general relativity.
Most accurate estimation on the mass of neutron stars has been done
in double neutron star binaries, and the neutron star masses in these
binaries are $\le 1.5 M_\odot$, which are consistent with rather soft EOS.
However, recently the neutron star mass in a white dwarf-neutron star
binary J1614$-$2230 has been estimated to be $(1.97\pm 0.04) M_\odot$
\cite{Dem10}.
This observation is very important because the mass estimation is
based on the detection of Shapiro delay, i.e., time delay of signals
due to the gravity of companion star. It is well known that Shapiro
delay is one of the key observation to have very good mass estimation.
If this number is confirmed by other observations, it can rule out many
soft EOS, including super-soft EOS. Hence the validity of soft symmetry
energy that gives soft EOS can be checked by future investigations
on the neutron star masses.

\section*{Acknowledgments}
This research was supported by the Daegu University Research Grant, 2010.

\end{document}